# Evaluating regulatory reform of network industries: a survey of empirical models based on categorical proxies

Andrea Bastianin
University of Milan

Paolo Castelnovo
University of Milan

Massimo Florio
University of Milan

This version: September 25, 2018

## Abstract

Proxies for regulatory reforms based on categorical variables are increasingly used in empirical evaluation models. We surveyed 63 studies that rely on such indices to analyze the effects of entry liberalization, privatization, unbundling, and independent regulation of the electricity, natural gas, and telecommunications sectors. We highlight methodological issues related to the use of these proxies. Next, taking stock of the literature, we provide practical advice for the design of the empirical strategy and discuss the selection of control and instrumental variables to attenuate endogeneity problems undermining identification of the effects of regulatory reforms.

Corresponding author: Massimo Florio, Department of Economics, Management and Quantitative Methods (DEMM), University of Milan, via Conservatorio 7, I-20122 Milan, Italy. E-mail: massimo.florio@unimi.it

## 1. Introduction

At least since the 1980s, governments around the world have implemented wide regulatory reforms that have reshaped network industries such as energy, telecommunications, and transport. The empirical evaluation of the societal impacts of these reforms[1] is essential to guiding policymakers and regulators in the selection of the most appropriate measures. This process seems straightforward: reforms are desirable when they yield economic and social benefits that outweigh their costs (Coglianese, 2012). While this simple description is backed-up by a well-established theoretical literature in public economics (see Boadway, 2012 for a survey), empirical assessments of regulatory reforms by means of econometric analyses are complex (Jamasb and Pollit, 2001). One of the main difficulties is how to build accurate empirical proxies for regulatory reforms and use them to identify their causal effects on key economic outcomes (Arndt and Oman, 2006; Knack, 2006). The recent controversy over the World Bank's competitiveness rankings, brought to the media's attention by former Chief Economist Paul Romer, exemplifies well the kind of criticisms to which extant proxies are subjected (Zumbrun and Talley, 2018).

This paper summarizes these criticalities focusing on entry liberalization, privatization, unbundling, and independent regulation of the electricity, natural gas and telecommunications sectors. We surveyed 63 empirical analyses that rely on dichotomous or categorical variables to evaluate the effects of regulatory reforms across sectors, countries, and over time. These proxies for reform, such as the OECD "Indicators of regulation in energy, transport and communications" (ETCR) or the "ICT Regulatory Tracker", recently released by the International Telecommunication Union, are based on a process that combines statistical data and subjective information from different sources (e.g., from surveys of business or experts). We discuss several issues involved in the measurement and assessment of regulatory reforms in network industries using categorical variables; next, taking stock of the literature, we provide recommendations to help researchers and policymakers avoid methodological pitfalls and errors in interpreting empirical results. First, since categorical variables summarize a variety of data, aggregation biases can be reduced relying, as far as possible, on the most disaggregated index. Second, given that conceptual errors involved in the construction of reform proxies are to some extent inevitable, re-coding categorical indices into dichotomous variables might mitigate the impact this issue. Third, we recommend using alternative proxies as robustness checks of the results and to facilitate cross-country comparisons. Fourth, properly handling the dynamics of the econometric specification is essential for capturing the forward-looking behavior of agents and to accurately describe the lags associated with the political process that leads to the implementation of reforms. Fifth, we review how to select appropriate control factors and valid instrumental variables to attenuate endogeneity problems that might undermine the identification of the effects of regulatory reforms.

---

[1] We use the term regulatory reforms to encompass any regulatory policy aiming to enhance the life of citizens and business; that is, any measure implemented by governments through legislative and non-legislative acts. In the literature, regulatory reforms are also referred to as structural reforms. See Section 2 for a more detailed discussion.

Empirical studies of the effects of regulatory reforms can be divided into two distinct groups depending on whether the reform proxy enters the econometric specification as a dependent variable or as an explanatory variable. The first class of models, which we do not analyze, is representative of the political economy literature on the historical determinants, success, and failure of reforms (see Obinger et al., 2016; Duso and Seldeslachts, 2010; Guerriero, 2013; Belloc et al., 2014, among many others). We focus on the strand of the literature where reform proxies are fed into regression models as explanatory variables with the aim of estimating their effects on various economic outcomes, ranging from prices to customers' satisfaction. Studies that are related to our paper are Nicoletti and Pryor (2006), Jamasb et al. (2017), Parker and Kirkpatric (2012), and Pollit (2009a, 2009b). We depart from previous surveys along two lines. First, we do not focus on economic outcomes (i.e., dependent variables), but mostly on regulatory reform indicators as explanatory variables. Second, we do not concentrate on a single sector or country, but on four different reforms: entry liberalization, privatization, unbundling, and the establishment of an independent regulatory authority in the electricity, natural gas and telecommunications sectors. In this way, we provide some guidelines for practitioners and academic researchers.

The rest of the paper is organized as follows. Section 2 sets the conceptual framework and describes the difficulties of translating general definitions of regulatory reforms into empirical proxies; Section 3 discusses potential pitfalls and methodological issues related to the use of categorical proxies for regulatory reforms; Section 4 surveys the literature focusing on the measurement of different regulatory reforms with categorical proxies; Section 5 discusses how to selected appropriate control factors and valid instrumental variables to attenuate endogeneity problems; Section 6 concludes.

## 2. Taking the theory of regulatory reforms to the data

OECD (2012, p. 5) stated that regulatory policy aims at "(…) *achieving government's objectives through the use of regulations, laws, and other instruments to deliver better economic and social outcomes and thus enhance the life of citizens and business*." This definition is broad enough to encompass most of the regulatory reforms implemented in network industries, including liberalization, privatization, unbundling, and the establishment of independent regulatory bodies. The OECD's definition also shows that theoretical models belonging to the "Ramsey-Samuelson-Guesnerie" tradition[2] can hardly be used to analyze them. In fact, within this framework, regulatory reforms are expected to change a "vector of signals", defined as variables affecting the behavior and welfare of individuals and firms, such as prices, rations, taxes, transfers, and shareholding rights (see e.g., Diamond and Mirrlees, 1971; Drèze and Stern, 1990, Gruber and Saez, 2002; Saez et al., 2012; Barrell and Weale, 2009; Kosonen, 2015; Laubach, 2009). This framework is too narrow to represent a proper theoretical background for the array of reforms subsumed in the OECD's definition of regulatory policy.

In fact, regulatory reforms in network industries do not necessarily involve a marginal

---

change to a vector of signals, such as tax rates or prices, nor an instantaneous variation in the social welfare function. Rather, they often imply a modification of the existing policy framework and are implemented with legislative packages that encompass a variety of instruments, such as primary laws, secondary regulations, subordinate rules, standards, administrative guidance and circulars, with complex interactions[3] (OECD, 2010; Goldberg, 1976).

While presenting a complete list of regulatory reforms is neither feasible, nor particularly informative, providing definitions of the four measures we analyze is useful to better illustrate the topics of the paper and the difficulties that emerge when these definitions need to be translated into quantitative or qualitative variables.

### 2.1 Regulatory reforms in network industries: definitions

We focus on liberalization, privatization, unbundling and the introduction of an independent regulatory body in electricity, natural gas and telecommunications sectors. Providing consistent definitions is not trivial in that these regulatory reforms are intertwined and their success depends on several factors including the order and the way in which they are implemented in different countries and sectors (e.g., Zhang et al., 2005).

Finding a precise definition of "liberalization" is problematic. In fact, this term is used to encompass several measures aimed at spurring competition. For instance, privatization and the unbundling of the network core facilities are often viewed as part of the liberalization process (see e.g., Erdogdu, 2013; Pompei, 2013; Nepal et al., 2016). To avoid definitional problems, we adopt a narrow definition and equate the term liberalization to "entry liberalization", defined as the removal of barriers to entry (i.e., any factors hindering entry in a market and hence representing an obstacle to competition). Privatization is more easily defined either as the transfer of property rights from the public to the private sector or as the participation of the private sector in the management of public assets (Finger and Künneke, 2011), in this paper we focus on the first aspect only, the change of utility ownership.

---

[3] In the case of the European Union, regulation of network services involves the adoption of both legislative and non-legislative acts, the so-called "soft law". Legislative acts (secondary law) comprise directives, regulations and decisions. Non-legislative acts include communications, green papers and white papers. "Soft laws" provide a correct interpretation of the primary and secondary laws. See Maresca (2013) and references therein.

We exclude from liberalization those obstacles faced by the incumbent firms when they entered the market in the first place, such as the sunk costs of building the network's core facilities. This exclusion allows for defining the concept of "unbundling" more precisely, although it remains intertwined with that of entry liberalization. The unbundling of network infrastructures aims at fostering competition in sectors where highly integrated firms operate (see Joskow, 1997). Integrated firms provide multiple services to exploit economies of scope that can derive from vertical integration (i.e., along different stages of the supply chain) or horizontal integration (i.e., the case of a multi-utility operating at a single stage of the production chain, but in different segments or sectors). Unbundling is thus the separation of the operation of network infrastructures from the stages of production and provision of services.[4] A key dimension of structural reform is the degree of separation, from accounting to ownership separation. See Baldwin et al. (2013) for a discussion.

Although liberalization, privatization, and unbundling have the common goal of fostering competition, their success hinges on the existence of an authority supervising and enforcing the rules necessary for their implementation. The establishment of an independent regulator is the fourth regulatory reform analyzed in our study. An independent regulatory authority (IRA) is an agency that independently from the interests of firms and of policymakers and other stakeholders can issue binding decisions and impose penalties, so to enforce the rules necessary for the actual implementation of reforms (see e.g., Armstrong and Sappington 2006; Bortolotti et al. 2013; Pollit. 2009a; Larsen et al. 2006). The importance of the IRA for the effectiveness of entry liberalization, privatization, and unbundling crucially depends on the actual power that the political system delegates to it, and also by cultural norms, political and legal traditions, and other institutional features. Therefore, while capturing legal or *de jure* independence is quite easily achieved looking at the statute of the regulatory authority, measuring *de facto* independence is at least as important, but more involved (see Bortolotti et al. 2013; Edwards and Waverman, 2006). In both cases, the measurement process and the resulting indicators share the same features of those used for other regulatory proxies, notably the possibility of being expressed as categorical variables based on the aggregation of different sources of information. Furthermore, in the European Union, a related regulatory reform involves the institution of supra-national regulators and their interactions with national regulatory bodies. The degree of centralization of regulatory policies at supra-national level is another example of a factor that could be measured with categorical variables and that is intricately related to the effectiveness of other regulatory reforms.[5]

---

[4] An alternative view adopted by some scholars is to reserve the term "unbundling" for measures aimed at regulating utilities that operate at the same level of the production chain, and use the term "restructuring" for the vertical separation of functions (e.g., generation, transmission and distribution).

## 2.2 Databases for use in measuring regulatory reforms

Given the complexity and interconnectedness of regulatory reforms, their quantitative assessment necessitates creating or resorting to artificial indicators that have no natural units of measure. A wide array of such indicators, which in most cases are categorical variables, have been used by scholars for their analyses (see e.g., Erdogdu, 2011a; Prati et al., 2013; Opolska, 2017; Koo et al., 2012; Grajek and Röller, 2012; Howard and Mazaheri, 2009; Nicoletti and Pryor, 2006). Data in these analyses have been compiled by international organizations, independent think tanks, and individual researchers.

In this section, we discuss four prominent databases (namely, the ETCR, the MOM, the EURI, and the ICT Tracker) that are representative of what is available to researchers focusing on network industries. We stress that this list is not exhaustive[6] and that our selection criteria were (*i*) their representativeness of the kind of empirical proxies for regulatory reforms discussed in this paper and (*ii*) the fact that they are widely used in the literature (see, among many others, Alesina et al., 2005; Brau et al., 2010; Belloc, 2013; Bacchiocchi et. al 2015; Faccio and Zingales, 2017; Copenhagen Economics 2005a,b; Edwards and Waverman, 2006; Montoya and Trillas, 2007; Ugur, 2009). All in all, the following discussion highlights the complexity of quantitative measurement and the trade-offs between the closeness to the underlying theoretical definitions of reform and data availability.

---

[5] Similarly, in the US the interaction between Federal and State level regulators is amenable to representation with such categorical variables. There is a substantial literature on decentralization or "fiscal federalism" (Bardhan, 2002 and Oates, 1999, for a survey). As far as network industries are concerned, we could not find any empirical studies dealing with the institution of supra-national regulators and their interactions with national regulatory bodies, nor a database focusing on this issue. A theoretical contribution related to this topic is Kumkar (2002) who analyzes the centralization of the regulation of the electricity markets within the European Union. For this reason, we do not cover this issue in our analysis.

*The OECD indicators of regulation in energy, transport, and communications (ETCR).* The ETCR database collects information about regulatory structures and policies for OECD and some non-OECD countries. ETCR indices have been widely used (e.g., Alesina et al., 2005; Brau et al., 2010; Belloc, 2013, among many others); however, we note that the database does not provide any information about the independence of the regulatory agencies. Therefore, in cases when this reform is of interest, ETCR data must be complemented with information from other sources.

Information is collected with a questionnaire sent to governments and complemented with publicly available data to create annual time series starting in the mid-1970s and regularly updated. Questionnaires are made of closed questions that can either be answered with numerical values (e.g., the market share of the largest company in the sector), or by selecting from a set of pre-specified answers (e.g., a question that can be answered with "yes" or "no", such as whether unbundling of the local loop is required). Qualitative information is coded into quantitative measures and then all answers are normalized in a range from zero to six, where values near zero indicate fewer restrictions to competition.[7] As shown in Figure 1, the ETCR index aggregates with equal weights indices for seven network industries: gas, electricity, air, rail, road transport, post and telecommunications. The structure of the ETCR database has evolved over time; here we refer to its latest release (Koske et al., 2015). For each sector, there are up to four sub-indices that cover different dimensions of the reforms: entry regulation, vertical separation, public ownership and market structure. One of the strengths of the ETCR database is that indicators are available for a very long horizon, making it suitable for panel-data analyses, see Figure 2.

---

[6] Another example is the dataset compiled by the IBM Global Business Services and used to build the Rail Liberalizations Index (LIB Index), which provides information on the progress of entry liberalization in the European rail transport markets.

Although there are sector-specific items, some of the questions in the ETCR database can be easily linked with the concepts of entry liberalization (i.e., entry regulation), privatization (i.e., public ownership), and unbundling (i.e., vertical separation).[8] As for electricity and natural gas, the ETCR database includes indices measuring the extent of privatization, entry liberalization and unbundling. As for telecommunications, only the first two regulatory reforms are tracked. For the natural gas sector, Figure 1 highlights that while the entry liberalization index can take on integer values from zero to six, the answers to the underlying questions have at most three alternatives or, in some cases, are based on re-scaled percentages. As we can see, the three questions in the top panel of Figure 1 are concerned with the quantification of barriers to entry. In the case of electricity, the first question is the same, but then there is also a second item about the existence of a liberalized wholesale market and a third item that investigates under which conditions consumers are free to choose their electricity supplier. In this case, we notice that only the first and third questions are tightly linked with the notion of barriers to entry. For telecommunications, there are three questions that can be answered choosing among three alternatives and that focus on the existence of barriers to entry in different segments of the market (i.e., trunk, international, and mobile). Privatization is investigated focusing only on the percentage of the government's ownership, while nothing is said about public vs. private management. Moreover, in the case of the electricity sector, there is a single question that does not discriminate about different segments (e.g., generation, transmission and distribution are all in the same question). The possible answers represent either re-scaled percentages or arbitrary codifications of percentages or qualitative items (e.g., "private" set to 0, "mostly private" set to 1.5, "mixed" equal to 3, and so on). Unbundling and the degree of vertical separation are measured with similar techniques.

---

[7] Consider the question: "What is the market share of the largest company in the sector?" The methodology assigns a score of zero if the share is smaller than 50%, three if it is between 50% and 90% and six if it is greater than 90%.

[8] We notice that items belonging to the "market structure" section can be either the result of regulatory reforms or elements used to track them. For this reason and since the underlying questions do not clearly represent any of the four regulatory reforms we study, from now on we disregard this part of the ETCR database.

*Market Opening Milestones (MOM) database*. The MOM database was developed by Copenhagen Economics (2005a, 2005b) for the European Commission and was used among others by Brau et al. (2010) and Ugur (2009). It collects information on several regulatory reforms in electricity, gas, telecommunications, transport, and postal service implemented by the EU-15 Member States over the 1990-2004 period. Since then it has not been further updated. The database tracks a variety of regulatory reform, referred to as Market Opening Milestones (MOMs), whose nature and number varies across sectors, but that can be easily associated with those we are interested in. For instance, in the case of the electricity sector, Copenhagen Economics focuses on the portion of the retail market that is open for free choice, third-party access regimes for the distribution and transmission networks, ownership of generation, and the extent of unbundling of the transmission and the distribution networks. As for the natural gas sector, the MOMs under scrutiny are the degree of unbundling of transmission and distribution system operators, ownership of supply companies, third-party access to the transmission grid, distribution grid and storage capacity, and degree of free choice of supplier.[9] For the telecommunications sector, relevant MOMs are the ownership of networks, third-party access regimes, and the extent of customer choice.

Indices in the MOM database have a zero value if the regulatory reform has yet to be implemented and a positive score, greater than zero and at most equal to one, after it has been implemented. The score assignment is based on expert estimates of the importance of the MOM for market opening and as such is highly subjective. The information collected in the MOM database is used to produce an overall indicator, the Market Opening Index, which tracks each sector's progress towards market opening. Despite its name, the Market Opening Index does not relate uniquely with entry liberalization; in fact, its components can be used to construct ETCR-like proxies for entry liberalization, privatization, and unbundling.[10] Therefore the Market Opening Index is closer in spirit to the ETCR broad index for a given sector. One key difference is the aggregation scheme, which in this case does not use equal weights, but involves a highly subjective process based on expert evaluation of the relative importance of different MOM. Like its underlying components also the aggregate index goes from zero to one where, in the parlance of Copenhagen Economics, a unit value denotes "full market opening". See Figure 3.

---

[9] Additional MOMs for the electricity sector are: tariff structure and congestion mechanisms wholesale market model. Further MOMs for the natural gas sector are: tariff structure in transmission pricing and regulation of end-user price

*The European Union Regulatory Institutions (EURI) database*. The EURI database and the associated EURI Independence Index, due to Edwards (2004) and used by Montoya and Trillas (2007) and Edwards and Waverman (2006), focus on regulatory independence in the telecommunications sectors of 15 EU Member States over the 1997-2003 period. The underlying data are sourced from reports of the European Commission and of the OECD. The aim of these indicators is to gauge the extent to which the IRA is independent of the country's government. Different aspects of regulatory independence are investigated focusing on powers, jurisdiction, composition, funding mechanisms, availability of resources, and other key features of the statute of NRAs. These data are collected for twelve items[11] and are summarized with dichotomous or categorical variables rescaled to take on at most four (integer or non-integer) values in the unit interval. See Figure 4.

---

[10] Interestingly, the MOM database allows for the reconstruction of indicators that in some cases are similar to those in the ETCR database. An entry liberalization index is obtained by summing the MOMs referring to third-party access and to the degree of free choice of supplier. A proxy for privatization can be produced focusing on the network ownership MOM. A correlation analysis of the indices in the MOM and ETCR databases was provided in the working paper version of Brau et al. (2010) who showed that the two set of indices are significantly correlated.

Building on these data, the EURI Independence Index is an unweighted sum of the twelve mentioned items (Edwards and Waverman, 2006). The resulting index ranges continuously from 0 to 12, with higher values corresponding to greater regulatory independence. Besides issues associated with the construction of the index, the EURI database has other limitations: it focuses only on a single sector, it is not regularly updated, and it does not collect information about the interaction between supra-national authorities and IRAs. Another limitation is that the EURI Independence index captures only *de jure* independence, while it tells nothing about factors contributing to *de facto* independence.

*The ICT Regulatory Tracker (ICTRT)*. The ICTRT, recently released by the International Telecommunication Union, covers about 190 countries over the 2007-2016 period. It relies on quantitative and qualitative information to build a set of 50 indicators that are grouped into four "clusters". See Figure 5. The first and second clusters (10 and 15 indicators, respectively) are concerned with regulatory independence and summarize the main features and the mandate of the regulatory authority.[12] The third cluster (11 indicators) encompasses a broader set of regulatory issues that are beyond the scope of our analysis; lastly, the 14 indices belonging to the fourth cluster track the level of competition in the ICT sector and as such are more related to the effects of specific measures, rather than with the measurement of the phase of the reform process.

---

[11] These are: whether the IRA is (i) single or multi-sector; (ii) a single or multi-member body; (iii) funded by government appropriations or industry fees and consumer levies; (iv) reports only to the executive government or also to the legislature; (v) has adequate powers regarding interconnection issues; (vi) shares its regulatory functions with the government; (vii) whether the legislature is involved in IRA member appointment; (viii) whether IRA member terms of appointment are guaranteed, allowing them to exercise regulatory power without concern for political factors that might influence their continued tenure; (ix) whether IRA member terms are renewable; whether IRA (x) staff and (xi) resources are adequate; and (xii) whether the IRA has been in operation for at least 2 years since its establishment.

[12] Examples of indicators belonging to the first cluster include variables accounting for the existence of separate telecommunications/ICT regulator; autonomy in decision-making; enforcement power; and a requirement for public consultations before decisions. Indicators in the second cluster ("Regulatory Mandate") provide information about who is in charge of regulating, for example, the service quality, licensing, interconnection rates and price regulation, radio frequency allocation and assignment, broadcasting and internet content and so on.

Each of the 50 indicators is given a score of 0, 1, or 2 and then they are aggregated with equal weights within each cluster to build four broad indices, which are in turn added up (again, with equal weights) to yield an overall indicator. This overall index ranges between 0 and 100, where higher values correspond to a more advanced regulatory environment. Countries are then classified into four categories: those with regulated public monopolies ($0 \leq$ score $< 40$); those that have implemented basic reform ($40 \leq$ score $< 70$); those enabling investment, innovation and access, stimulating competition in service and content delivery, and consumer protection ($70 \leq$ score $< 85$); and those with integrated regulation ($85 \leq$ score $\leq 100$). These definitions and aggregation process are discretionary and identify both the effect and the status of the reform. Lastly, similarly to the EURI Index, the ICT Tracker focuses only on the telecommunications sector and on *de jure* independence, but it has the advantage of being constantly updated and covering a larger sample of countries. Recent studies relying on this database include Faccio and Zingales (2017), Ortiz et al. (2017) and Mailu and Waema (2016).

## 3. Issues related to categorical proxies for regulatory reforms

As highlighted in the overview of databases, most regulatory reform proxies share two common features: they are compiled by transforming quantitative and qualitative information into normalized scores and are aggregated with a bottom-up approach. If score assignment and aggregation are meaningful and consistent over time and across countries, the effects and intensity of reforms could be easily tracked, as illustrated in Figures 2 and 3. However, building such scoring systems and aggregating them with bottom-up approaches involves several issues that can be grouped into three main categories: (*i*) conceptual errors, (*ii*) discretization, and (*iii*) aggregation errors. These three issues cause a broader class of empirical problems that goes under the header of *measurement error*. This fourth issue is related with the discussion in Section 5, where we analyze how researchers have dealt with this and other factors that can cause endogeneity problems and hence undermine the identification of the effects of regulatory reforms.

## 3.1    Conceptual errors

Conceptual errors arise when the regulatory reform proxy does not clearly relate to a narrow theoretical description of the measure the researcher is interested to study. Moreover, the analysis is prone to interpretational and identification problems when the variable designed as a reform indicator might also be the outcome of some macroeconomic shock. For instance, the market share of the largest electricity generator is part of the policy and can be used as an empirical proxy when the regulator forces the incumbent to divest generation capacity. By comparison, if the market share has changed in response to an exogenous technological shock that altered the optimal production plan, not only for the incumbent but also for other firms the sector, the validity of the reform proxy is questionable. In this case, the technological shock is a confounding factor that causes an omitted variable bias and hence ultimately affects the evidence in favor or against the reform. Another way of introducing a conceptual error is to rely on regulatory reform indicators that are not able to fully capture the actual extent of a regulatory reform. A case in point is the distinction between *de jure* and *de facto* regulatory independence.

## 3.2    Scale discretization

Most regulatory reform proxies, such as those in the ETCR or MOM database, have a multidimensional and often discrete scale. The reason for relying on discretized variables is the lack of a common unit of measure. Discretization has many drawbacks (Rucker et al., 2015). First, potentially useful variability is discarded and second, this loss of variability reduces the precision of in-sample and out-of-sample predictions. When the reform process is summarized with a dichotomous variable, there might be a substantial reduction in statistical power, which increases the chance of both Type-I (false positive) and Type-II error (false negative). Moreover, Lien and Balakrishnan (2005) showed that the dichotomization of the independent variable reduces goodness-of-fit and may increase or decrease the regression slope. Dichotomization also leads to interpretational issues. In fact, using a dummy variable within a panel regression to measure the existence of barriers to entry, implies that market opening would be of the same magnitude in all countries. Lastly, discretization may be arbitrary, because the data provide no guidance for defining the scoring systems.

## 3.3    Aggregation

No valid economic criteria are usually available to guide the aggregation of different categorical variables using bottom-up approaches. With no *a priori* information on the relative importance of each proxy, equal weights are often assigned to different aspects of reforms and upper-level proxies are simple averages or sums of the lower-level variables. This may bias estimates, affect inference, and influence the evaluation of reforms since aggregate scores depend on arbitrarily selected equal weights. Besides these problems, the use of aggregate reform measures in place of sub-indices leads to a loss of information because the effects resulting from distinct aspects of the regulation are not separately identified.

## 3.4    Measurement error

Measurement error (ME) identifies any deviation from the true value of a variable that arises in the definitional or measurement stage (Asher, 1974) and therefore subsumes all the issues discussed so far. Mis-measurement is likely to be particularly severe in the case of regulatory reform proxies since they rely on an array of non-standard data sources such as media coverage, legal and non-legal acts, and surveys that are then processed by expert judgment. In this case, misreporting by subjects, coding and other errors are likely to inflate ME.[13] Moreover, even in the presence of correctly measured variables, what we called "*conceptual errors*" are another source of ME. In fact, observed data often do not correspond to the exact theoretical concept the analyst is interested in. Another relevant issue comes from the fact that long time series are typically more precisely measured in the present than in the past. A case in point is the ETCR database, as its extension backward in time is expected to create a systematic measurement error, given that the quality of data is probably correlated with time. In fact, the reconstruction of data and smaller coverage for the early part of the sample makes data in most recent years more accurate than in the past. We turn now to a discussion of the empirical literature on regulatory reforms having in mind the issues identified above.

## 4. Categorical proxies for regulatory reforms: a survey of the literature

Earlier literature on the impact of regulatory reforms on network industries considered their impact on a wide array of dependent variables. Consumer prices are the most straightforward, important and widely used proxy for welfare changes induced by the regulatory reforms (Price and Hancock, 1998, for the theoretical background; Bacchiocchi et al. 2011; Bacchiocchi et al. 2015; Brau et al. 2010; Erdogdu 2011b; Estache et. al. 2006; Fiorio and Florio, 2013; Growitsch and Stronzik, 2014; Hattori and Tsutsui, 2004; Hyland, 2016; Nagayama, 2007, 2009; Steiner, 2001; Wallsten, 2001, among others, for empirical analyses). An alternative approach is to rely on consumers' satisfaction surveys to measure perceived welfare changes (see e.g., Clifton et al., 2010, 2011, 2014; Bacchiocchi et al. 2011; Ferrari et al., 2011; Fiorio et al. 2013).

Further dependent variables considered in this strand of the literature are aggregate output

---

[13] In the benchmark case of classical ME, least squares estimates will be downward biased (or "attenuated") and inconsistent. In the literature on regulatory reforms, since indicators are often categorical or dummy variables and models can also be non-linear, classical ME is rarely a good benchmark. See the surveys by Angrist and Krueger (1999), Bound et al. (2001), and Hausman (2001) for a comprehensive discussion.

(e.g., Copenhagen Economics, 2005a; Prati et al., 2013), productivity (e.g., Pompei, 2013; Borghi et al. 2016), investment (e.g., Alesina et al. 2005; Gugler et al. 2013; Nardi, 2012; Grajek and Roller, 2012; Cambini and Rondi, 2010), quality of service (e.g., Zhang et al., 2005, 2008; Polemis, 2016; Polemis and Stengos, 2017; Koo et al. 2012; Briglauer, 2014, 2015) and emissions of pollutants (e.g., Asane-Otoo, 2016; Clò et al., 2017).

Our aim here is not to summarize the conclusions of the literature on the merits of the regulatory reforms, but rather to highlight if and how researchers have acknowledged the existence of measurement and interpretational issues. For this reason, our analysis, which includes studies published between 2001 and 2018, does not focus on a specific timeframe, set of countries or on a given economic outcome.

## 4.1 Aggregate indices of regulatory reforms

Instead of analyzing exclusively specific measures, many studies rely on indicators that aggregate different regulatory reforms.[14] A case in point is Alesina et al. (2005), who used the aggregate ETCR index. This indicator collapses in a single number all the information relating with privatization, unbundling and entry liberalization. Similarly, Ugur (2009) and Copenhagen Economics (2005*a,b*) relied on the aggregate Market Opening Index sourced from the MOM database. While this approach provides a summary of the impact of regulatory reforms in network industries, it does not allow for distinguishing the measures that drive the estimated effect and is clearly subject to aggregation biases due to the use of equal or subjective weights. Relying on equal weights implicitly means that all regulatory reforms move a given outcome in the same direction and with the same magnitude. Moreover, time-invariant weights convey the idea that neither the phase of the reform process (e.g., legal or ownership unbundling) nor the sequencing of different measures matter. Lastly, some of the studies that focus on aggregate reform indicators equate them to all-encompassing terms such as "deregulation" or "liberalization" (see e.g., Erdogdu, 2013; Pompei, 2013; Nepal et al., 2016). Terms that are vague or not tightly linked to theoretical definitions, or that lack theoretical definition at all, confound the interpretation of results.

---

[14] See, among many others, Alesina (2005); Bacchiocchi et al. (2011); Bacchiocchi et al. (2015); Erdogdu (2011a, b, 2013); Fiorio and Florio (2013); Grajek and Roller (2012); Polemis and Stengos (2017); Pompei (2013); Prati et al. (2013); Zhang et al. (2008); Copenhagen Economics (2005); Ugur (2009).

## 4.2   Entry liberalization

The extent to which barriers to entry have been removed is measured by aggregating information on different issues, such as the terms and conditions of third-party access and the share of the retail market that is open to consumer choice.[15] Aggregate indicators of entry liberalization might introduce a bias in their estimated effects for two reasons. First, different aspects of this regulatory reform are aggregated with subjective weights and, second, this aggregation process induces a loss of information. Both problems can be either attenuated, by relying on data-reduction techniques or completely avoided using only lower-level indicators. Data-reduction techniques, such as Principal Component Analysis, have been successfully implemented to calculate weights and synthetic reform proxies in many settings (e.g., Ferrari and Manzi, 2014; Poggi and Florio, 2010). While these methods may yield a weighting scheme, they do not subsume any economic theory, leave almost no control to the analyst, and are severely affected by the presence of outlying observations. An alternative is to check the robustness of results to different weighting schemes with randomly drawing weights (Koske et al., 2015). Yet another option, implemented by Fiorio and Florio (2013) and Steiner (2001), is to include the items composing the ETCR entry liberalization index directly in their empirical specifications to disentangle their relative importance. When the aim is to study the effects of entry liberalizations on customer satisfaction (Bacchiocchi et al., 2011) or energy deprivation (Poggi and Florio, 2010), researchers estimate non-linear models, such as probit or logit specifications. When moving from linear to non-linear models the effects of classical or non-

---

[15] Studies using proxies for entry liberalizations are, among the others, Bacchiocchi et al. (2015); Brau et al. (2010); Fiorio and Florio (2013); Growitsch and Stronzik (2014); Hattori and Tsutsui (2004); Steiner (2001); Alesina et al. (2005); Asane-Otoo (2016); Botasso and Conti (2010); Gugler et al. (2013); Kim et al. (2012); Pompei (2013); Polemis (2016).

classical ME are harder to evaluate and to address.[16]

### 4.3 Privatization

Variables relating with the transfer of shares of ownership from the public to the private sector have been made available in several databases or constructed by individual researchers. Privatization indices are built starting from a continuous variable bounded in the 0-100 interval, where we interpret 100 to mean that all the shares are owned by the government. Scholars source these variables from different databases including the ETCR[17] or directly construct them from company ownership data (e.g., Borghi et al. 2016; Bortolotti et al. 2013; Opolska, 2017). These proxies are then used either directly (e.g., Alesina et al., 2005; Bacchiocchi et al., 2011; Growitsch and Stronzik 2014; Hattori and Tsutsui, 2004; Steiner, 2001) or after transforming them into a dummy variable (e.g., Bacchiocchi et al. 2015; Asane-Otoo, 2016; Howard and Mazaheri 2009; Zhang et al. 2005). While this transformation is in principle legitimate and is done to carry out before-after comparisons (e.g., to identify price changes due to the privatization) it has some drawbacks. A dummy does not allow for measuring the extent and magnitude of privatization and measurement error (classical or not) in the underlying ownership variable would introduce non-classical measurement error in the dummy variable. In fact, a dichotomous variable can only be misclassified in one of two directions: a true "zero" classified as "one" or a true "one" classified as "zero". In a bivariate linear regression setting, a mis-measured dummy variable still leads to an attenuation bias, but generalizations of these results to other settings with more than one regressor is not possible[18] (Card, 1996). A related problem is the inclusion of squared ownership variables to capture nonlinear relations between privatizations and the outcome variable of interest. Polynomial transformations inflate the size of the attenuation bias coming from classical measurement errors (Griliches and Ringstad, 1970). Alesina et al. (2005) used the square of the ETCR privatization index, as well as the squares of more aggregate regulatory reform proxies, to model their nonlinear effects on investment. Erdogdu (2011a) followed a similar approach to study the impact of regulatory reforms, proxied by a composite reform index, in the electricity sector.

---

[16] In fact, while instrumental variable estimation often provides a solution to the ME problem in linear models, it does not deliver consistent estimates in a nonlinear regression framework. See Bound et al. (2001), Hausman (2001) and Schennach (2016) for a survey of ME in nonlinear models.

[17] See, among the others, Bacchiocchi et al. (2011); Bacchiocchi et al. (2015); Brau et al. (2010); Fiorio and Florio (2013); Growitsch and Stronzik (2014); Polemis (2016); Alesina et al. (2005); Asane-Otoo (2016); Belloc et al. (2013); Clò et al. (2017); Pompei (2013).

[18] A way to deal with the issue of misclassification of the reform variable is to rely on the bound approach of Bollinger (1996).

The measurement of privatizations, much like that of any other regulatory reforms, is also subject to aggregation biases. The ETCR database, as well as the data used in other papers, such as Hyland (2016) and Nagayama (2007, 2009), are sometimes available at a different level of aggregation. For the natural gas sector, the ETCR database provides ownership information about the largest firm in the production, transmission and distribution segments. In this case, using the disaggregated indices could reduce aggregation biases related to the discretionary approach of relying on equal weights. An alternative solution would be to collect ownership data for all firms operating in different sectors. These data can then be used directly in empirical analyses as in Borghi et al. (2016) or combined with other information, such as the firms' share of sector revenue, to construct sector indices based on time-varying aggregation weights. Such an aggregation scheme would put more weight on firms with larger market shares.

## 4.4 Unbundling

The degree of separation of the operation of network infrastructures from the stages of production and provision of services is typically measured with a categorical variable with an ordering going from full vertical integration, to complete unbundling. Since this is a relatively simplistic, but consistent way of identifying the sequence of legislative acts that triggers the unbundling process, categorical variables representative of this process have been used in several papers, including Bacchiocchi et al. (2015); Brau et al. (2010); Fiorio and Florio (2012); Growitsch and Stronzik (2014); Hattori and Tsutsui (2004); Hyland (2016); Steiner (2001); Asane-Otoo (2016); Gugler et al. (2013); Nardi (2012); Nepal et al. (2016); Pompei (2013). In this case, the arbitrary discretization of the phases of the vertical separation process (e.g., the ETCR database assigns three scores: 0 to ownership separation, 3 to legal and/or accounting separation and 6 fully integrated segments of the natural gas sector) is likely to generate further issues because ordinal data are treated as cardinal variables. To reduce the distortions that this cardinalization might introduce, researchers have proposed two approaches. As suggested by Fiorio and Florio (2011), one might determine with extensive robustness checks whether results critically depend on the cardinalization adopted. Alternatively, Bacchiocchi et al. (2015), Growitsch and Stronzik (2014), Zhang et al. (2005) and Nagayama (2007), among others, suggested collapsing scores into a dichotomous variable, or aggregating them in very wide brackets. There is a clear trade-off between statistical variability and discretization issues: the dichotomization dramatically reduces the cardinalization error, but also leads to information losses. Moreover, the use of dummy variables is potentially subject to other issues. In the case of a semi-logarithmic model, the percentage change in $y$ due to a discrete change in $D$ from 0 to 1 is in fact given by $p = 100 \times (\exp\{c\} - 1)$; however, using the Ordinary Least Squares (OLS) estimate of $c$, denoted as $\hat{c}$, yields a biased estimator for $p$. A better solution is to rely on $\hat{p} = 100 \times [exp\{\hat{c} - 0.5\ \hat{V}(\hat{c})\}\ \text{-}1\ ]$, where $\hat{V}(\hat{c})$ is the OLS estimate of the variance of $\hat{c}$ (Kennedy, 1981). Among the surveyed studies, only Nepal et al. (2016), analyzing the unbundling of the New Zealand electricity market, addressed this issue as suggested.

Lastly, the study of unbundling might also be affected by the level of aggregation of the reform indicator. For instance, any aggregation bias might be attenuated considering disaggregated indices (see e.g., Hyland, 2016) in place of an aggregate unbundling indicator that does not distinguish between generation, transmission, distribution, and the retail sale of electricity.

## 4.5   Independent Regulation

Papers focusing on this regulatory reform are rooted in the strand of the literature dealing with central bank independence that also relies on similar categorical variables to measure the autonomy of the monetary policy authority (see Acemoglu et al., 2008; Alesina and Summers, 1993; Cukierman, 1992; Grilli et al. 1991, among many others). Studies investigating the impact of the establishment of an independent regulatory authority are based either on dichotomous variables, that simply measure the existence of a separate regulator (e.g., Fink et al. 2003; Wallsten 2001, 2002; Gutierrez and Berg, 2000; Ros 2003; Zhang et al. 2005; 2008; Nagayama 2007; Erdogdu 2011a; Estache et al. 2002, 2006; Bortolotti et al. 2013), or on more sophisticated aggregated indicators (e.g., Gutierrez, 2003; Edwards and Waverman, 2006; Bauer, 2003), that include also other characteristics of the statute and mandate of the regulators (e.g., funding sources and financial autonomy, the ability of the government to overrule the regulators' decisions). The introduction of IRAs, much like any reforms of regulatory institutions, is expected to have complex effects on the effectiveness of other measures enforced to enhance competition in network industries. In regression models, indices measuring the independence of regulator are often interacted with proxies for entry liberalizations (e.g., Polemis, 2016), unbundling (e.g., Nagayama, 2007; Erdogdu 2011a) and privatizations (e.g., Bauer, 2003; Edwards and Waverman, 2006; Fink et al., 2002; Wallsten, 2001; Estache et al., 2002; Bortolotti et al., 2013). A related topic is that of the sequencing of different regulatory reforms. For instance, Zhang et al. (2005) and Wallsten (2002) showed that whether an IRA existed or not before privatization is key for their ability to enhance competition. Lastly, institutional quality might also affect the performance of independent regulatory authorities. This topic has been dealt with in different ways. Estache et al. (2006) interacted the regulatory independence indicator with a corruption and an investment risk index that are used as indices of the quality of institutions. Alternatively, Bortolotti et al. (2013) included proxies for the quality of institutions directly among controls, without interactions with the regulatory independence index. Lastly, Prati et al. (2013) split their sample of countries into quartiles based on the quality of their institutional environment, measured by the strength of constraints on the executive power and the risk of expropriation.

## 5. Identifying the effects of regulatory reforms

From an econometric standpoint, the empirical analyses reviewed in the previous section can all be framed in the setting of the generalized linear (GLM):

$$g[E(y \mid \mathbf{x})] = \mathbf{x}'\boldsymbol{\beta} \qquad y \sim LEF \qquad (1)$$

where $y$ is the outcome of interest, which can be a cross-sectional, time-series or panel variable, $\mathbf{x}$ denotes the set of explanatory variables including also the reform proxy, $\boldsymbol{\beta}$ is the vector of unknown parameters, $g(.)$ is called *link function* and *LEF* indicates a density belonging to a linear exponential family. Examples of *LEF* densities are the Bernoulli distribution for binary data, the Poisson and negative binomial distributions for count data, and the normal distribution for continuous data (Gourieroux et al., 1984). Combining different link functions and *LEF* densities yields a wide array of models that encompass all the specifications commonly used in the econometric literature about regulatory reforms in network industries[19] (McCullagh and Nelder, 1989). Two classes of models are commonly used in this literature. Linear models, when the dependent variable is continuous, such as log-prices, log-productivity and GDP growth (see e.g., Hyland 2016, Hattori and Tsutsui, 2014; Prati et al., 2013, among many others). Models for limited dependent variables, when the outcome variable is a dummy, such as consumers' satisfaction or energy deprivation (see e.g., Bacchiocchi et al., 2011; Ferrari et al., 2011; Fiorio and Florio, 2011; Clifton et al., 2011, 2014; Poggi and Florio, 2010).

Within the GLM framework the interpretation of estimated coefficients, from a purely statistical point of view, is relatively straightforward. However, economic interpretation requires more caution. Different issues may indeed undermine the identification of the causal effects of regulatory reforms on the dependent variable of interest. Measurement error, already discussed in Sections 3 and 4, is just one of the causes of endogeneity problems, that might also arise because of omitted variables and simultaneity. In this section, we briefly review how researchers have addressed the identification problem. We focus on the selection of appropriate control factors and valid instrumental variables in panel-data analyses that represent the vast majority of the studies we survey. Both these choices are critical to take into account possible endogeneity problems.

---

[19] For illustrative purposes, let $\mu \equiv E(y \mid \mathbf{x})$ and $\eta \equiv g(\mu)$. The role of the link function $g(.)$ is to map $E(y \mid \mathbf{x})$ to $\eta = \mathbf{x}'\boldsymbol{\beta}$. When $g(.)$ is the identity function and y is a continuous normally distributed variable, such as (the log of) the electricity price, we get the linear regression model: $E(y \mid \mathbf{x}) = \mathbf{x}'\boldsymbol{\beta}$. When $g(.)$ is the logit function (inverse of the standard Normal cumulative distribution) and $y$ is a binary variable with Bernoulli distribution, such as consumers' satisfaction (e.g., $y = 1$ corresponds to "satisfied", $y = 0$ means "not satisfied"), we obtain the Logit (Probit) model. Moreover as already stated Equation (1) is suitable also for analyses based on panel data. Under the assumption that the link function $g(.)$ is the identity function and that the outcome variable is normally distributed, we get a linear one-way error component model for panel data: $y_{it} = \alpha + \mathbf{x}_{it}'\boldsymbol{\beta} + u_{it}$, for $i = 1,\ldots, N$ and $t = 1,\ldots,T$. Where $u_{it} = \mu_i + \upsilon_{it}$ is the error term, $\mu_i$ is the unobservable individual-specific effect that accounts for the heterogeneity among statistical units, $\upsilon_{it}$ is the remainder disturbance.

## 5.1    On the choice of control variables

Virtually all empirical specifications omit some relevant, but possibly unobservable explanatory variables. In these cases, the error term of the regression will pick up the influence of such confounding factors and hence omitted variable bias is likely. In the ideal case, both the outcome variable and the reform proxy are correctly measured. Omitting other relevant variables because they might be measured with error would simply substitute the bias due to measurement error with an "omitted variable bias". Moreover, McCallum (1972) showed that using a proxy for a latent variable will induce a smaller asymptotic bias than simply dropping the variable from the model. The proper treatment of unobserved heterogeneity or the inclusion of a suitable set of controls might help mitigate such bias. Either fixed or random effects are used to account for unobserved heterogeneity among statistical units (see e.g., Growitsch and Stronzik, 2014; Hattori and Tsustsui, 2004; Gugler et al., 2013; Erdogdu 2011a, 2013; Koo et al. 2012; Agiakloglou and Polemis, 2018). Although the statistical units that are analyzed obviously determine whether firm, individual, sector, or country fixed effects are included in the regression model, dummy variables accounting for heterogeneity across countries are included in virtually all of the studies we survey.

While country fixed effects capture a wide array of unobserved characteristics that are fixed in time, researchers sometimes include time-varying explanatory variables accounting for institutional features and the quality of the political system (i.e., measures of impartiality, bureaucratic quality, corruption), judicial and labour market indicators, and economic indicators measuring the involvement of the government in the economy (see Bortolotti et al., 2013; Grajek and Roller, 2012; Gugler et al., 2013; Koo et al. 2012; Wallsten, 2001). Two examples are the Fraser index of economic freedom (see e.g., Polemis, 2016; Polemis and Stengos, 2017; Zhang et al., 2018; Agiakloglou and Polemis, 2018; Gwartney et al., 2012) and measures of investors' protection (Bortolotti et al, 2013). Alternatively, interactions between regulatory reform proxies and indicators of institutional quality could be useful to understand how the effectiveness of reforms varies with countries' institutional quality (see e.g., Borghi et al., 2016; Estache et al., 2006).

Since economic relations are not static in nature, an omitted variable problem might arise also when the effect of lagged dependent variables is neglected or, more generally, in the presence of serial correlation. Panel data exhibit two sources of correlation over time. The first, unobserved heterogeneity, arises because the same individual is observed over multiple time periods. The second, true state dependence, occurs when correlation over time is due to the causal mechanism explaining the fact that a variable is determined by its own lagged values; leading examples are prices or aggregate investment (e.g., Brau et al. 2010, Alesina et al. 2005). In case of path dependence, the most straightforward way to introduce some dynamics is to rely on an autoregressive distributed lag model and employ a GMM estimator to face the endogeneity issues that may arise when including lagged values of the dependent variable among the regressors (see Prati et al., 2013; Alesina et al., 2005; Guegler et al., 2013; Pompei, 2013; Polemis, 2016; Grajek and Roller, 2012; Briglauer, 2015; Growitsch and Stronzik, 2014). However, including a lagged dependent variable may lead to underestimating some of the effects of the other explanatory variables (see Howard and Mazaheri, 2009).

Even static models may, at least implicitly, consider the time dimension. For instance, Steiner (2001) and Hattori and Tsutsui (2004), who focused on electricity prices, used variables that track years needed to complete liberalization and privatization processes to capture expectations of the impact of reforms on prices. An alternative is including variables counting the number of years during which reforms have entered into force (Howard and Mazaheri, 2009). The sequencing of reforms may also be relevant to determine their effectiveness. This is studied in Zhang et al. (2005), who included a set of dummy variables denoting whether competition was introduced before privatization and whether regulation was implemented before privatization of the electricity generation sector.

In addition to modeling unobserved heterogeneity with fixed or random effects and dealing with the dynamics, most analyses also include a set of variables to control for confounding factors determined by the nature of the dependent variable. For instance, when analyzing the effects of regulatory reforms on prices, natural candidates are demand- and supply-side controls. Demand-side factors include proxies for the size of the market (see Bacchiocchi et al. 2011; Hyland, 2016; Erdogdu, 2011b; Wallsten, 2001; Erdogdu, 2011b), the price and the demand for substitutes (see e.g., Growitsh and Stronzik, 2014 and Bacchiocchi et al., 2011). Supply-side drivers proxy the change in unit costs, that in turn are correlated with input prices and, in the longer-run, also with technological shifts. For example, in the case of electricity production, unit costs depend on the price of energy inputs such natural gas, coal, nuclear, hydropower, solar, and wind, as well as on production and distribution losses (see Bacchiocchi et al. 2015; Nagayama, 2007). Finally, including a time trend would be useful to capture technological progress (see Polemis 2016; Polemis and Stengos, 2017).[20]

---

[20] This list is clearly not exhaustive. For instance, when explaining investment levels it is crucial to account for the real interest rate to control for the cost of capital (see e.g., Alesina et al, 2005; Gugler et al. 2013; Cambini and Rondi, 2010); environmental outcomes are likely affected by the environmental regulation enforced (Clò et al, 2017; Asane-Otoo, 2016); the adoption of a given ICT technology can be influenced by consumers' characteristics (Briglauer, 2014; Howard and Mazaheri, 2009); among the determinants of R&D expenditure it is important to include firm-level characteristics (see Kim et al., 2012) and other macroeconomic variables (see Erdogdu, 2013). When dealing with customer satisfaction demographic characteristics are good candidates (see Clifton et al. 2011; Fiorio and Florio, 2011) are crucial determinants to be considered; the underlying objective economic variable (i.e., the actual price level), if available, would be also useful control variables (see Fiorio and Florio, 2011).

## 5.2 Finding valid instrumental variables

While not simple, the search for valid instruments is at the core of several research papers. In fact, regulatory reforms are often endogenously determined, since factors influencing changes in the outcome variable of interest may also affect the likelihood that policymakers implement them, as well as the timing and sequence in which they are enforced. For example, it may be the case that, at least initially, regulatory reforms lead to higher prices, or that countries with higher electricity prices are more likely to reform their electricity markets (see Hyland, 2016; Nagayama, 2009). The choice of giving up the control of state-owned enterprises to private investors may be determined either by the government need to raise money selling highly profitable public utilities (Cambini and Rondi, 2010) or by the willingness to sell off very inefficient and unprofitable entities. Therefore, the post-reform performance also depends on the pre-reform efficiency and profits that in turn have determined the government's decision to privatize. Similarly, when analyzing the effects of the establishment of an independent regulator, it should be kept in mind that governments might have an incentive to set up a regulatory agency in sectors where profitability is expected to be higher (Bortolotti et al., 2013). Even if several studies recognize the existence of this complex interplay between regulatory reforms and the outcome variable of interest, not all of them try to overcome the reverse causality problem (see e.g., Wallsten, 2001).

The simplest solution to avoid possible simultaneity biases is to lag the reform proxy as well as other explanatory variables that might lead to reverse causality issues (see Bacchiocchi et al. 2011; Bortolotti et al. 2013). A relatively more sophisticated approach is to make use of instrumental variables. Valid instrumental variables must be uncorrelated with the error term of the regression and must be correlated with the endogenous variable of interest. Given the difficulty of finding valid instruments, dynamic panel data techniques, such as the Arellano and Bond (1991) estimator are often used to deal with possible endogeneity problems. Lagged values of the dependent variable and of the regressors are used as instruments after first differencing the regression model (see Hyland, 2016; Alesina et al., 2005; Polemis, 2016).

Alternatively, political economy variables, such as the degree of political fragmentation, are usually considered valid instruments. Political institutions are likely to affect the reform process, such as the size of the stake retained in public utilities or the extent of government residual regulatory power (Bortolotti et al. 2013) but should be uncorrelated with the dependent variable and hence solve simultaneity issues. Studies using political economy variables as instruments for reform indicators are Nagayama (2009), Grajek and Roller (2012), Cambini and Rondi (2010, 2017), Gual and Trillas (2003), Grajek and Roller (2012) and Bortolotti et al. (2013). However, Edward and Waverman (2006) in their analysis of regulatory independence criticized the validity of these instruments, arguing that it is hard to claim they are uncorrelated with sector or firm performance. They relied on an alternative approach, originally developed by Evans and Kessides (1993), to identify a valid instrument for the index of regulatory independence whose exogeneity with respect to regulated interconnection rates paid by entrants to incumbents might be questionable. They created a regulatory reform index based on country rankings according to the EURI-I index. The expectation is that while the implementation of reforms enhancing regulatory independence will be affected by high interconnection rates, it is unlikely that changes in these rates will affect the ranking of countries.

## 6. Concluding remarks

We reviewed the use of categorical variables in econometric studies of regulatory reforms in selected network industries. This is the first time such a review has been provided in the empirical regulatory economics literature. We started from the distinction between the theory of reform in the Ramsey-Samuelson-Guesnerie tradition and the problem of evaluating regulatory reform when it cannot be read as a marginal change to a specific economic signal. This distinction clearly matters to empirical analysis as the assessment of reform is prone to both methodological and interpretational errors. The interpretation of estimated coefficients is also affected because reform is not usually a marginal change in the variable of interest, but rather a transition from one policy regime to another.

The bottom line of our analysis is that while moving from the theory to the analysis of reform might be challenging, empirical evaluation is encouraged. In fact, the quantitative measurement of how social welfare responds to a change of circumstances, such as those implied by a reform, is not only at the core of applied welfare economics, but also essential because evaluation of outcomes informs policymakers about both success and failure, and alerts them to the need for adjustments to regulation consistent with policy goals (OECD, 2010, p. 9). Put simply, "evidence on the outcomes of regulatory policies should help policymakers design regulatory measures that work better" (Parker and Kirkpatrick, 2012), Therefore, although extant proxies have shortcomings, they are nevertheless the best quantitative information available to researchers interested in empirically analyzing the impact of reforms and as such, they can be used to deepen our knowledge of the effectiveness of different economic policies. In this review, we have shown that while researchers have proposed different solutions to attenuate endogeneity problems that might undermine the identification of the effects of regulatory reforms, further research is needed to address the methodological issues we discussed. We have also offered our practical advice on how to deal with some of the unavoidable implications of collapsing regulatory reforms into proxies based on categorical variables derived from the available databases.

**Acknowledgments**: this version of the manuscript greatly benefited from the insightful comments of the Editor Janice Beecher and three anonymous reviewers. The authors gratefully acknowledge financial support from the Jean Monnet Network EUSERS (http://users.unimi.it/eusers/).

**Figure 1. Structure of the OECD's Energy, Transport and Communications Regulation (ETCR) database.**

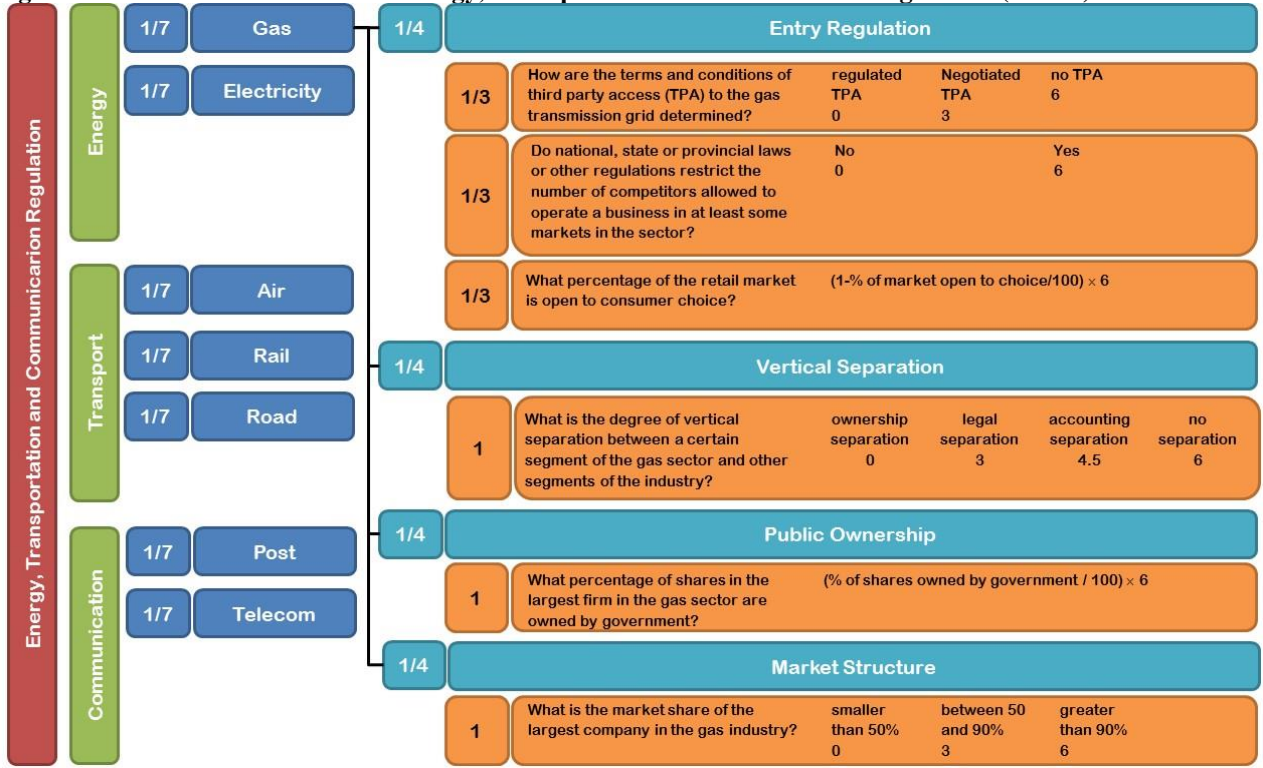

*Notes:* The underlying questionnaire for the gas sector is shown. Numbers are weights assigned to sectors, topics and questions and used for aggregation purposes.
*Source:* authors' elaboration using data in Koske et al. (2015).

**Figure 2. Trends of ECTR reform indicators in EU-15 countries for the natural gas sector, 1975 – 2013.**

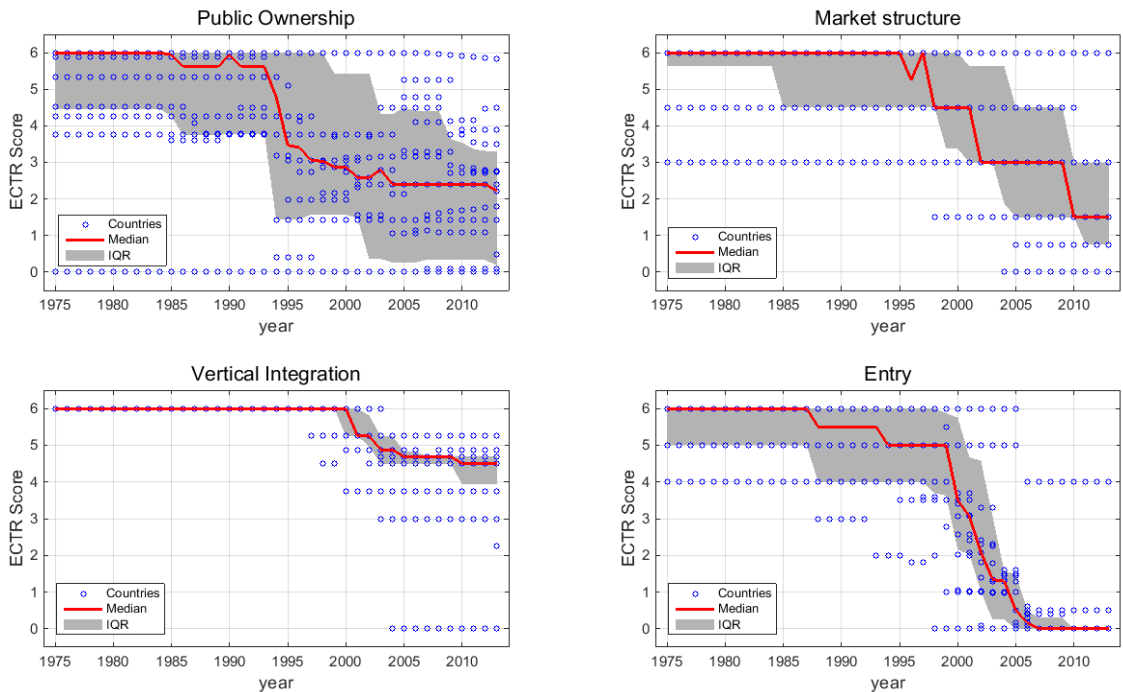

*Notes:* each panel shows one of the indicators of Energy, Transport and Communications Regulation (ETCR), for individual EU15 countries (dots), the median value of the indicator (line) and the interquartile range, (IQR, shaded area). ECTR reform indicators score from 0 to 6.
*Source:* authors' calculation on data sourced from the ETCR database (Koske et al., 2015).

**Figure 3. Market Opening Indices for the electricity, gas and telecommunications industries, 1990-2003**

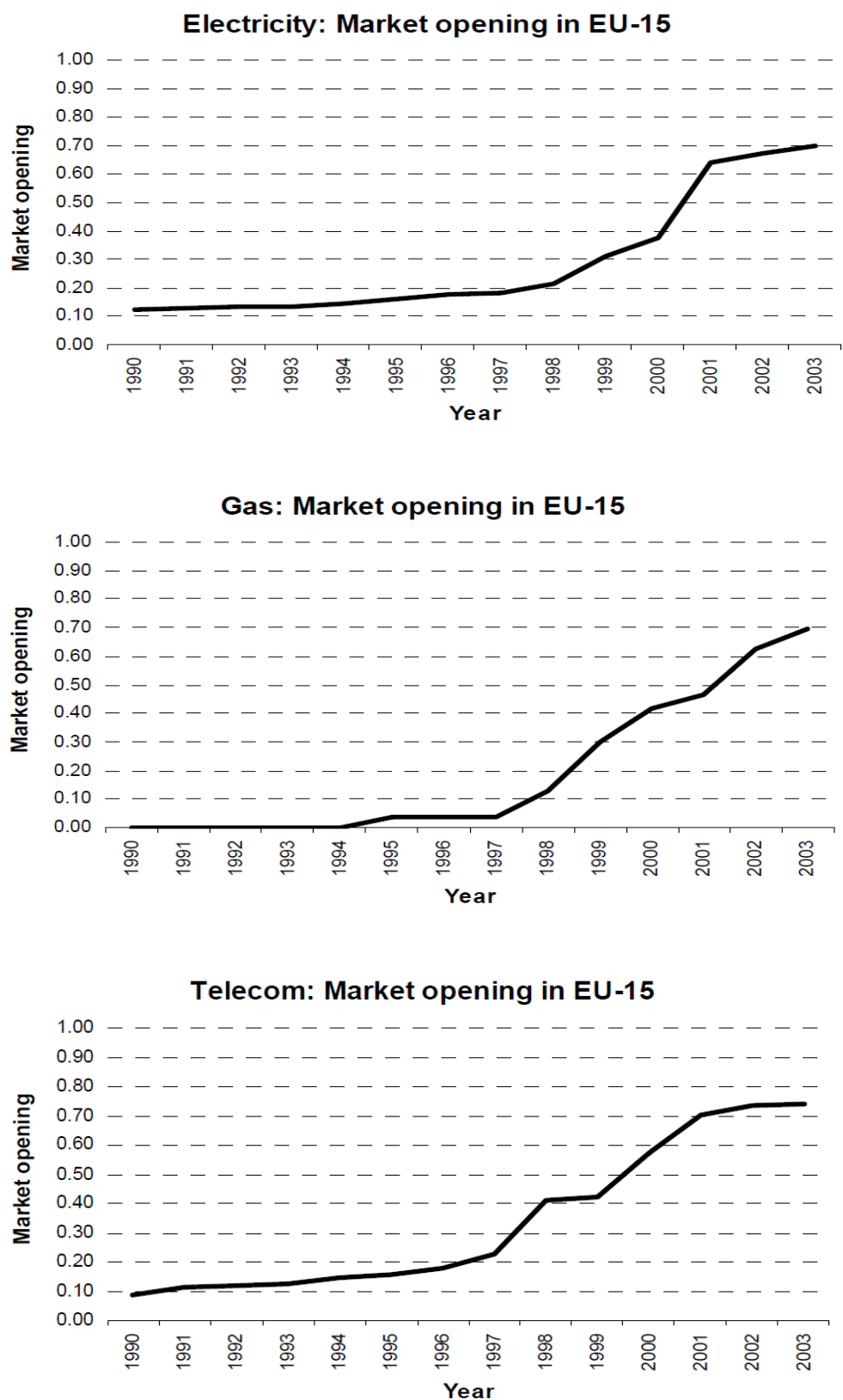

**Figure 4. European Union Regulatory Institutions - Independence (EURI-I) Index Elements (1998 and 2003)**

| Country | A | | B | | DK | | FIN | | F | | D | | EL | | IRL | | I | | L | | NL | | P | | E | | S | | UK | |
|---|---|---|---|---|---|---|---|---|---|---|---|---|---|---|---|---|---|---|---|---|---|---|---|---|---|---|---|---|---|---|
| Year | 98 | 03 | 98 | 03 | 98 | 03 | 98 | 03 | 98 | 03 | 98 | 03 | 98 | 03 | 98 | 03 | 98 | 03 | 98 | 03 | 98 | 03 | 98 | 03 | 98 | 03 | 98 | 03 | 98 | 03 |
| EURI-I | 1.5 | 5 | 5 | 4.5 | 5.75 | 5.75 | 5.5 | 4.5 | 6.5 | 6.5 | 7 | 8 | 5.5 | 7.5 | 5.5 | 8.5 | 8.5 | 10.25 | 4.5 | 5.5 | 5.75 | 7.75 | 7.75 | 9.5 | 6.75 | 6.75 | 7.75 | 8.75 | 5.75 | 5.75 |
| EURI-I (0-1 Scale) | 0.125 | 0.417 | 0.417 | 0.375 | 0.479 | 0.479 | 0.458 | 0.375 | 0.542 | 0.542 | 0.583 | 0.667 | 0.458 | 0.625 | 0.458 | 0.708 | 0.708 | 0.854 | 0.375 | 0.458 | 0.479 | 0.646 | 0.646 | 0.792 | 0.563 | 0.563 | 0.646 | 0.729 | 0.479 | 0.479 |
| Multi-sector | 0 | 1 | 1 | 1 | 0 | 0 | 0 | 0 | 0 | 0 | 0 | 0 | 0 | 1 | 1 | 1 | 1 | 1 | 1 | 1 | 1 | 1 | 1 | 1 | 0 | 0 | 1 | 1 | 0 | 0 |
| Multi-member | 0 | 0 | 0 | 0 | 0 | 0 | 0 | 0 | 1 | 1 | 1 | 1 | 1 | 1 | 0 | 1 | 1 | 1 | 1 | 1 | 1 | 1 | 1 | 1 | 1 | 1 | 1 | 1 | 0 | 0 |
| Funding | 1 | 1 | 1 | 1 | 0.75 | 0.75 | 1 | 1 | 0 | 0 | 0 | 0 | 1 | 1 | 1 | 1 | 0 | 0.25 | 1 | 1 | 0.75 | 0.75 | 0.75 | 1 | 0.75 | 0.75 | 0.75 | 0.75 | 0.75 | 0.75 |
| Reporting | 0 | 0 | 0 | 0 | 0 | 0 | 0.5 | 0.5 | 0.5 | 0.5 | 1 | 1 | 0 | 0 | 0.5 | 0.5 | 1 | 1 | 0.5 | 0.5 | 0 | 0 | 0 | 0.5 | 0 | 0 | 0 | 0 | 0.5 | 0.5 |
| Interconnect Powers | 0 | 1 | 0 | 0 | 1 | 1 | 1 | 0 | 1 | 0 | 0 | 0 | 0 | 1 | 1 | 1 | 1 | 1 | 0 | 0 | 0 | 1 | 0 | 1 | 1 | 1 | 0 | 1 | 1 | 1 |
| Shared Roles | 0 | 1 | 0 | 0 | 1 | 1 | 0 | 0 | 0 | 0 | 1 | 1 | 0 | 1 | 1 | 1 | 0 | 0 | 0 | 0 | 1 | 1 | 1 | 1 | 0 | 0 | 1 | 1 | 0 | 0 |
| Legislative Appointment | 0 | 0 | 0 | 0 | 0 | 0 | 0 | 0 | 1 | 1 | 1 | 1 | 1 | 1 | 0 | 0 | 1 | 1 | 0 | 0 | 0 | 0 | 0 | 0 | 1 | 1 | 0 | 0 | 0 | 0 |
| Fixed Terms | 0 | 0 | 0 | 0 | 0 | 0 | 0 | 0 | 1 | 1 | 1 | 1 | 1 | 1 | 1 | 1 | 1 | 1 | 1 | 1 | 1 | 1 | 1 | 1 | 1 | 1 | 1 | 1 | 1 | 1 |
| Renewable Terms | 0 | 0 | 0 | 0 | 0 | 0 | 0 | 0 | 1 | 1 | 0 | 0 | 0 | 0 | 0 | 0 | 1 | 1 | 0 | 0 | 0 | 0 | 0 | 0 | 0 | 0 | 0 | 0 | 0 | 0 |
| Staff | 0.5 | 0 | 1 | 0.5 | 1 | 1 | 1 | 1 | 0.5 | 0.5 | 1 | 1 | 0 | 0 | 0.5 | 0.5 | 0.5 | 1 | 0 | 0 | 0.5 | 0.5 | 1 | 1 | 0.5 | 0.5 | 1 | 1 | 1 | 1 |
| Budget | 0 | 0 | 1 | 1 | 1 | 1 | 1 | 1 | 0.5 | 0.5 | 1 | 1 | 0.5 | 0.5 | 0.5 | 0.5 | 1 | 1 | 0 | 0 | 0.5 | 0.5 | 1 | 1 | 0.5 | 0.5 | 1 | 1 | 0.5 | 0.5 |
| Experience | 0 | 1 | 1 | 1 | 1 | 1 | 1 | 1 | 0 | 1 | 0 | 1 | 1 | 1 | 0 | 1 | 1 | 0 | 1 | 0 | 0 | 1 | 1 | 1 | 1 | 1 | 1 | 1 | 1 | 1 |

*Notes: t*he figure shows the 12 institutional elements measured by the EURI Database. Each element is measured as either a categorical or dummy variable on a 0-1 scale. The EURI-I index represents the sum of these 12 measures. It can therefore range continuously from 0 to 12, with higher values corresponding to greater regulatory independence. Although the minimum and maximum in the sample considered by Edwards and Waverman (2006) are 1.5 and 10.25, respectively.
*Source*: Edwards and Waverman (2006), p.44.

**Figure 5. Structure of the ICT Regulatory Tracker Database**

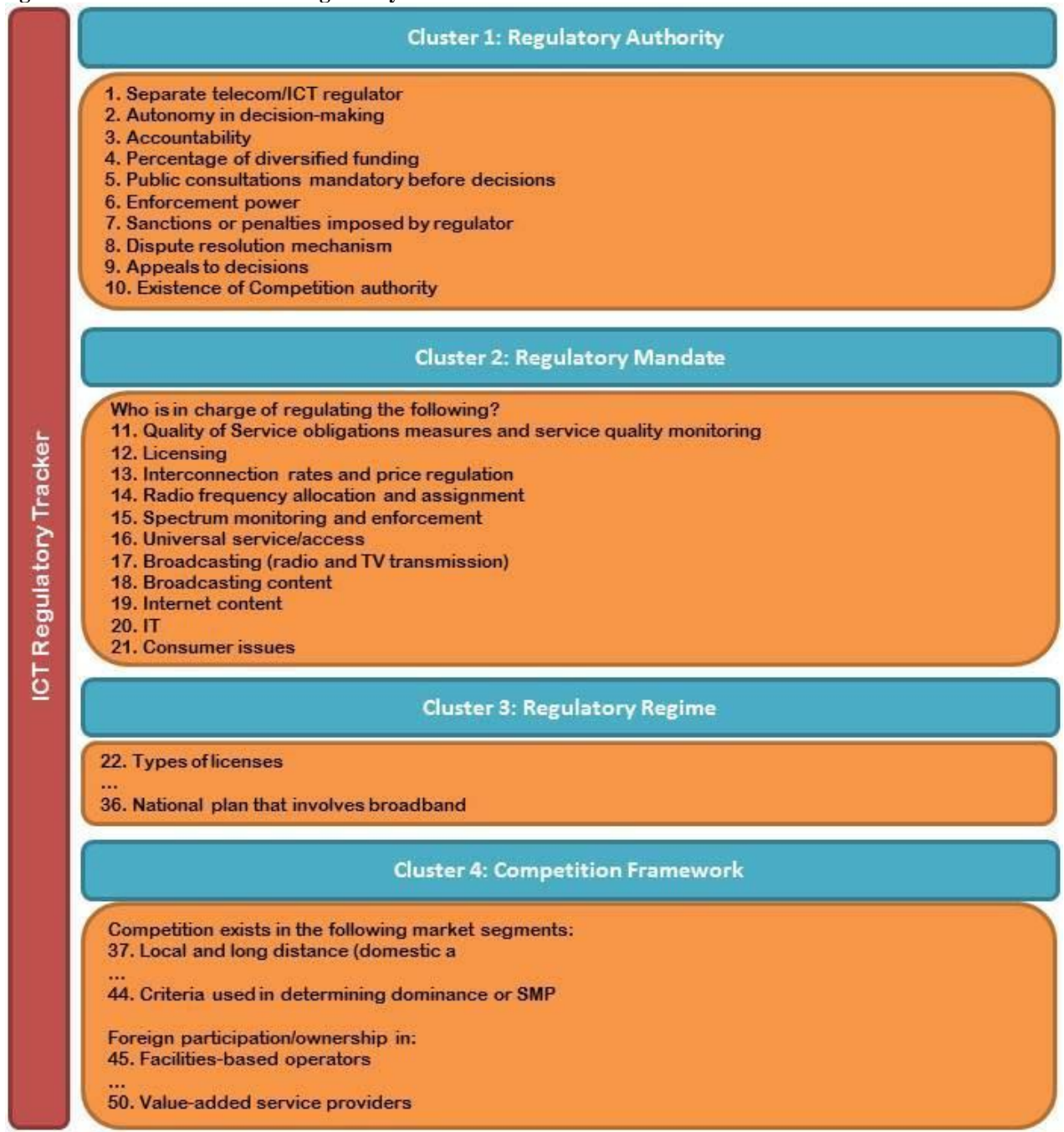

*Notes:* the ITC Regulatory Tracker index is the sum of the 50 questions shown in the figure with equal weights.
*Source:* authors' elaboration based on information available at https://www.itu.int